\documentclass[useAMS,usenatbib,usegraphicx]{mn2e}

\usepackage{amsmath}
\usepackage{amsfonts}
\usepackage{rotating}
\usepackage{color}
\usepackage{pdflscape}
\usepackage{txfonts}
\usepackage{natbib, twoopt}
\usepackage[flushleft]{threeparttable}
\usepackage{longtable}
\usepackage{supertabular,booktabs}
\bibpunct{(}{)}{;}{a}{}{,} %% natbib format like A&A and ApJ 
\newcommand{\teff}{$T_{\mbox{\scriptsize eff}}$}
\newcommand{\kms}{$\mbox{kms}^{-1}$}
\newcommand{\logg}{log\,$g$}

\title[Binary clusters NGC 5617 and Trumpler 22]{Binary open clusters in the Milky Way: photometric and spectroscopic analysis of NGC 5617 and Trumpler 22\thanks{Based on observations obtained at the Anglo-Australian Telescope, Siding Spring Observatory, Australia }} 

\author[De Silva et al.]{G.M. De Silva$^{1,2}$\thanks{E-mail: gayandhi.desilva@aao.gov.au}, G. Carraro$^{3,8}$, V. D'Orazi$^{4,5,7}$, V. Efremova$^{1}$,
H. Macpherson$^{1,5}$, S. Martell$^{6}$, \newauthor L. Rizzo$^{9}$\\
$^{1}$Australian Astronomical Observatory, 105 Delhi Rd, NSW 2113, Australia\\
$^{2}$Sydney Institute for Astronomy, School of Physics, The University of Sydney, NSW 2006, Australia\\
$^{3}$ESO, Alonso de Cordova 3107, 19001, Santiago de Chile, Chile\\
$^{4}$INAF - Osservatorio Astronomico di Padova, vicolo dell'Osservatorio 5, 35122, Padova, Italy\\
$^{5}$Monash Center for Astrophysics,School of Physics and Astronomy, Monash University, VIC 3800, Australia\\
$^{6}$School of Physics, University of New South Wales, Sydney, NSW 2052, Australia\\
$^{7}$Department of Physics and Astronomy, Macquarie University, North Ryde, NSW 2109, Australia\\
$^{8}$Dipartimento di Astronomia e Fisica, Universita di Padova, Vicolo dell'Osservatorio 3, I-35122, Padova, Italy\\
$^{9}$Facultad de Ciencias Astron\'omicas y Geof\'isicas (UNLP), Universidad de La Plata (CONICET, UNLP), Paseo del Bosque s/n, La Plata, Argentina\\
}

\begin{document}
%\date{Accepted 1988 December 15. Received 1988 December 14; in original form 1988 October 11}

%\pagerange{\pageref{firstpage}--\pageref{lastpage}} \pubyear{2002}

\maketitle

\label{firstpage}

\begin{abstract}
Using photometry and high resolution spectroscopy we investigate for the first time the physical connection between the open clusters NGC 5617 and Trumpler 22. Based on new CCD photometry we report their spatial proximity and common age of $\sim$70 Myr. Based on high-resolution spectra collected using the HERMES and UCLES spectrographs on the Anglo-Australian telescope, we present radial velocities and abundances for Fe, Na, Mg, Al, Si, Ca, and Ni. The measured radial velocities are -38.63 $\pm$ 2.25 \kms\ for NGC 5617 and -38.46 $\pm$ 2.08 \kms\ for Trumpler 22. The mean metallicity of NGC 5617 was found to be [Fe/H] = $-$0.18 $\pm$ 0.02 and for Trumpler 22 was found to be [Fe/H] = $-$0.17 $\pm$ 0.04. The two clusters share similar abundances across the other elements, indicative of a common chemical enrichment history of these clusters. Together with common motions and ages we confirm that NGC 5617 and Trumpler 22 are a primordial binary cluster pair in the Milky Way.

\end{abstract}
 

\section[]{Introduction}
\label{section:introduction}

The current model of the global star formation hierarchy is that giant molecular clouds fragment into cloud cores, producing stellar complexes, OB associations and open clusters which dissipate into individual stars \citep{efremov95}. The exact mechanism that lead from the contraction of molecular clouds to star clusters is not completely understood, and likely there are different paths that lead to the formation of different systems of clusters. In the latter stages of this hierarchy, it has been suggested that star clusters could form in pairs or multiples \citep{marcos2009}. We refer to such clusters as primordial binary clusters.

Primordial binary clusters are expected to be transient in nature, where possible subsequent evolutionary paths include merging of the pair, tidal disruption of one or both clusters, and separation into two bound clusters. The models of \cite{bhatia90} suggest that binary cluster lifetimes range from a few 10$^{6}$ yrs to 4x10$^{7}$ yrs depending on factors such as cluster separation, tidal force of the parental galaxy and encounters with giant molecular clouds. The identification of primordial binary clusters is further complicated with possible other explanations for their origins. The mechanisms of tidal capture and resonant trapping can place two unrelated clusters in pairs \citep{Leon1999}, and these tend to eventually merge \citep{deOliveira2000a,deOliveira2000b}. Some candidate binary clusters could be merely optical doubles due to super-positioning along the field of view. 

In the Large and Small Magellanic Clouds (LMC and SMC), at least 10\% of the known open cluster systems may be in pairs and perhaps more than 50\% of these are primordial binary clusters \citep{bhatia1988,dieball2000,dieball2002}. Interaction between the LMC and the SMC may be the conditions giving rise to the formation of primordial binary star clusters \citep{fujimoto}. Binary and multiple star clusters have been seen in other violent environments including in the Antennae galaxies \citep{fall05,whitmore05} and in the young starburst galaxy M51 \citep{larsen,bastian}, presumably initiated by galaxyÐgalaxy interactions. The merger of large binary cluster systems have been suggested to lead to the formation of massive globular clusters, such as Omega Cen \citep{minniti2004} and NGC 1851 \citep{1851}.

In the Milky Way ~10\% of the Galactic open clusters have been proposed to be in binary or multiple systems \cite[e.g.][]{marcos2010,sub95}. The most well established binary cluster in the Galaxy is the pair $h$ and $\chi$ Persei \citep{uribe,dufton,marco}. None of the other proposed binary candidates have been verified with spectroscopic studies.  \cite{kopchev08} studied the binary candidate pair NGC 7031/NGC 7086 and based on photometry, found a discrepancy in age, which may be due to large age uncertainties, and encouraged radial velocity follow-up. \cite{vazquez2010} searched for candidate binary clusters in the 3rd Galactic Quadrant with a negative outcome and concluded binary clusters would form preferentially in environments closer to the environment of the LMC, suggesting further investigations in denser and more violent regions of the Milky Way, such as the inner Galaxy.

The apparent lack of confirmed binary or multiple open clusters in the Galaxy leads to several questions. Did the Milky Way not undergo a violent past to form such structures? If it did, perhaps the timescales have been too short to preserve any binary clusters today, as the transient nature of binary clusters meant they are no longer in existence as a binary system. Or have we not looked hard enough for binary clusters in our own Galaxy? 

A possible tracer of cluster origins is the motion and chemical abundance composition of their stars. Presumably if two clusters formed at about the same time and site, then they should have similar ages, motions and chemical properties, assuming that the larger gas cloud was sufficiently mixed, with no major contamination events (e.g. supernovae) taking place during the cluster formation timescales. As the chemical abundance patterns in low-mass stars are preserved during their lives, they reflect the conditions of their birth site. Hence a pair of primordial binaries should share a common age, radial velocity and chemical composition. Binary clusters caused by tidal capture may show common motion, but the chemical properties and ages are likely to be different. For any optical doubles we expect both the motion and chemical composition to be different between the two clusters.

In this paper we present the first spectroscopic study of stars in the binary open cluster pair Trumpler 22 \& NGC 5617. They are a strong primordial binary candidate with estimated ages of 70$\pm$10Myr (this study) and the pair are spatially separated by only ~20 pc \citep{sub95}. Stellar membership has been explored in the literature for NGC 5617 \citep[e.g.][]{Carraro04,Ahumada, frinchaboy, Mermilliod,Orsatti,Carraro2011}, however very few studies have targeted Trumpler 22.
This paper is organised as follows: Section \ref{sec:phot} presents the photometric data on the target cluster which also formed the selection criteria for the subsequent spectroscopic observations. In Section \ref{sec:rv}  we present the radial velocity analysis. In Section \ref{sec:chem}  we discuss the spectroscopic analysis. The findings are summarised in Section \ref{sec:summary} .

\section{Photometry}\label{sec:phot}

The photometry used in this paper comes from two sources. In the case of NGC~5617, modern CCD photometry in the UBVI passbands is available from \cite{Carraro2011}. In this study, estimates of the cluster fundamental parameters are derived. The age is constrained to 70$\pm$10 Myr, while the distance from the Sun is 2.1$\pm$0.2 kpc , and the reddening
E(B-V)=0.45$\pm$0.05. According to this study, the reddening law in the line of sight is normal ($R_V$=3.1).

\begin{center}
\begin{figure*}
\includegraphics[width=15cm]{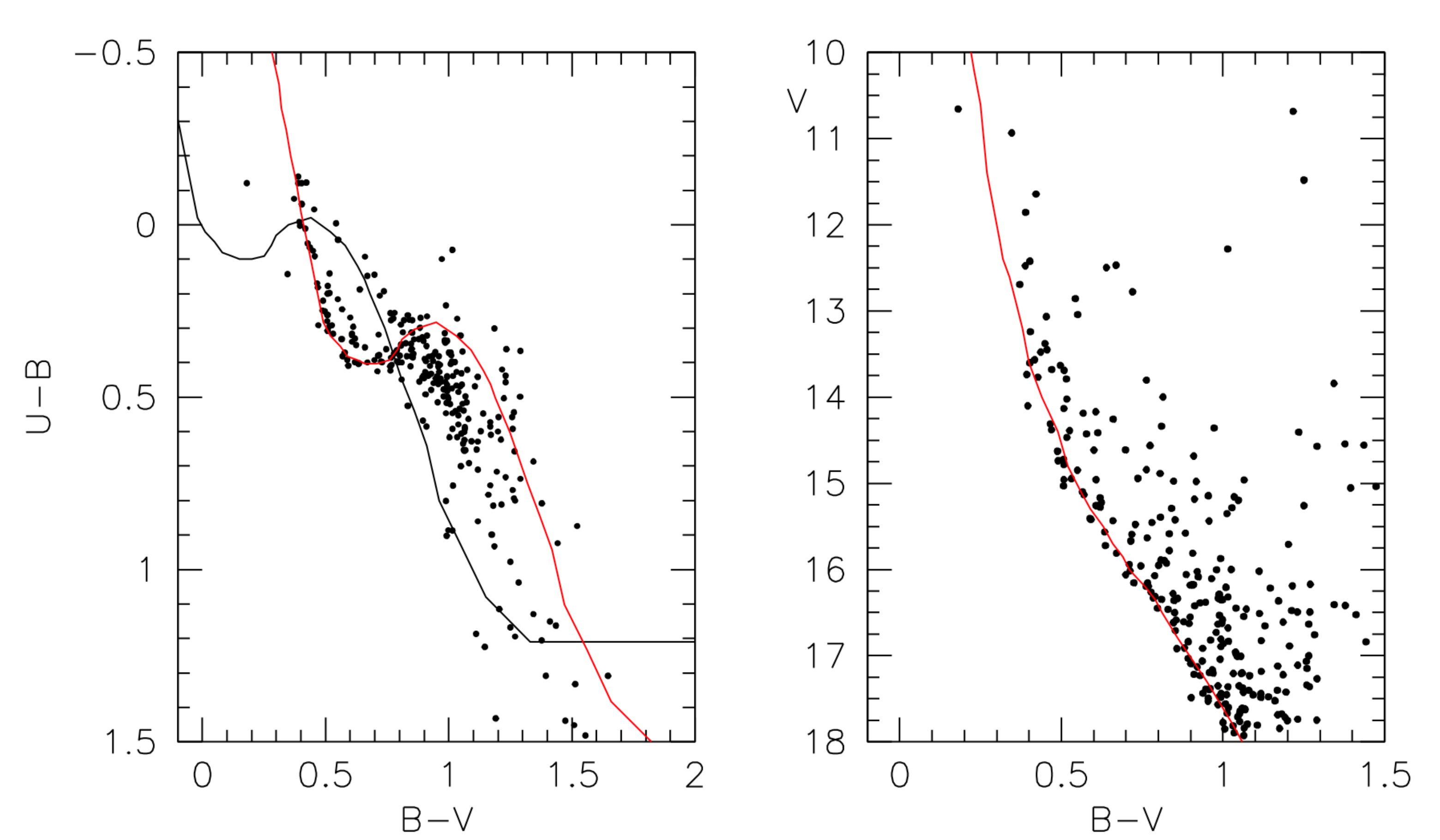}
\caption{{\bf Left panel:} Color-color diagram for Trumpler~22 stars within 4 arcmin from the cluster nominal center. The solid lines are zero-age main sequence relations for no-reddening (black line) and for E(B-V)=0.48 (red line). This latter line nicely fit the bulk of early type stars. {\bf Right panel:} Color-magnitude diagram for the same stars as in the left panel. The solid line has been displaced horizontally , by E(B-V)=0.48, and, vertically, by (m-M)=13.10.}
\label{f:phot1}
\end{figure*}
\end{center}

\begin{table}
\tabcolsep 0.1truecm
\caption{$UBVI$ photometric observations of Trumpler ~22 and Landolt standard stars on  June 3, 2011.}
\begin{tabular}{cccc}
\hline
\noalign{\smallskip}
Field & Filter & Exposures (s) & airmass (X)\\
\noalign{\smallskip}
\hline\hline
\noalign{\smallskip}
\hline
                                    Trumpler 22   & \textit{U}  & 30, 2x180, 900     & 1.07$-$1.28\\
                                                          & \textit{B}  &10, 2x120,600       & 1.07 $-$1.27\\
                                                         & \textit{V}   & 5, 2x60, 300        & 1.07$-$1.27\\
                                                         & \textit{I}    & 5, 2x60, 300        & 1.07$-$1.26\\
                                   PG1323        & \textit{U}  & 4x240                   & 1.17$-$2.05\\
                                                        & \textit{B}  & 4x120                   &  1.18$-$2.00\\
                                                        & \textit{V}  &  4x60                    &  1.18$-$2.05\\
                                                        & \textit{I}    & 4x60                    &  1.20$-$2.10\\
                                   PG16333     & \textit{U}  & 4x240                    & 1.17$-$1.87\\
                                                       & \textit{B}  & 4x120                    &  1.18$-$1.95\\
                                                       & \textit{V}  &  4x60                     &  1.18$-$1.92\\
                                                       & \textit{I}    & 4x60                     &  1.20$-$1.88\\
                                   Mark   A      & \textit{U}  & 4x240                    & 1.03$-$1.57\\
                                                       & \textit{B}  & 4x120                    &  1.03$-$1.55\\
                                                       & \textit{V}  &  4x60                     &  1.03$-$1.54\\
                                                       & \textit{I}    & 4x60                     &  1.03$-$1.56\\
\hline
\noalign{\smallskip}
\hline
\end{tabular}
\end{table}
 
For Trumpler 22, only old photographic data exist \citep{Haug78}, and for this reason we exploit here a new data-set. 
Photometry in UBVI was acquired at Las Campanas Observatory (LCO) on the nights from June 03, 2011 and are published here for the first time. We used the SITe$\#$3 CCD detector onboard the Swope 1.0m telescope\footnote{http://www.lco.cl/telescopes-information/henrietta-swope/}. With a pixel scale of 0.435 arcsec/pixel, this CCD allows to cover 14.8 $\times$ 22.8 arcmin on sky.  We stress that this setup (telescope/instrument) is the same that \cite{Carraro2011} used for NGC~5617.
The night was photometric with seeing ranging from 0.8 to 1.5 arcsec. We obtained multiple exposures per filter, and observed three times along the night the standard star field PG~1323, Mark~A, and  and PG~1633 , to cover a wide airmass range, and to secure proper photometric calibration (see Table~1). 
After removing problematic stars, and stars having only a few
observations in Landolt's (1992) catalog, our photometric solution
for the run was extracted from 
a grand total of 63 measurements per filter, and turned out to be:\\

\noindent
$ U = u + (5.004\pm0.010) + (0.49\pm0.010) \times X + (0.129\pm0.013) \times (U-B)$ \\
$ B = b + (3.283\pm0.006) + (0.25\pm0.010) \times X + (0.040\pm0.008) \times (B-V)$ \\
$ V = v + (3.204\pm0.005) + (0.16\pm0.010) \times X - (0.066\pm0.008) \times (B-V)$ \\
$ I = i + (3.508\pm0.005) + (0.08\pm0.010) \times X + (0.037\pm0.006) \times (V-I)$ \\
\noindent
where $X$ indicates the airmass.\\

\noindent
After standard pre-processing, photometry was extracted using the DAOPHOT/ALLFRAME package \citep{stetson}. The photometry data-set has finally astrometrized using the 2MASS\footnote{http://www.ipac.caltech.edu/2mass/} catalog. The data will be made available at the SIMBAD Astronomical Database\footnote{http://simbad.u-strasbg.fr/simbad/}.\

We compared our photometry with \citep{Haug78} photographic photometry for 65 stars in common. The comparison reads:
$\delta$V =$-$0.05$\pm$0.12, $\delta$(B-V) =$-$0.09$\pm$0.11, and $\delta$(U-B) = $-$0.11$\pm$0.20, in the direction of our photometry minus \citep{Haug78}. Such systematic differences are not worrisome,
and can be explained by small zero-points differences caused by the use of different telescopes, and different techniques (variable PSF vs fixed aperture diaphgram).

In Fig\ref{f:phot1},  (left panel), we show  the color-color diagram (CCD), that we used to estimate the cluster reddening, while in the right panel we shown Trumpler~22 color magnitude diagram (CMD), used to estimate the distance. Only stars within 4 arcmin from the cluster nominal centre are plotted. The reddening is estimated by displacing an empirical  zero age main sequence \citep[ZAMS,][]{schmidt-kaler} along the reddening line. The un-reddened ZAMS is shown as a black line, while the red line is the same ZAMS, shifted by E(B-V)=0.48$\pm$0.08. The sequence of early type stars is tight enough to assume we are facing a real cluster.
In the right panel the same sample of stars is fitted with a ZAMS for the same reddening and for an apparent distance modulus (m-M)=13.1$\pm$0.2.  Both reddening, distance modulus, and the associated uncertainties
have been estimated via the usual visual inspection method.

This implies a distance of 2.10$\pm$0.3 kpc, and we conclude that Trumpler~22 and NGC~5617 share the same heliocentric distance.

\begin{center}
\begin{figure}
\includegraphics[width=10cm]{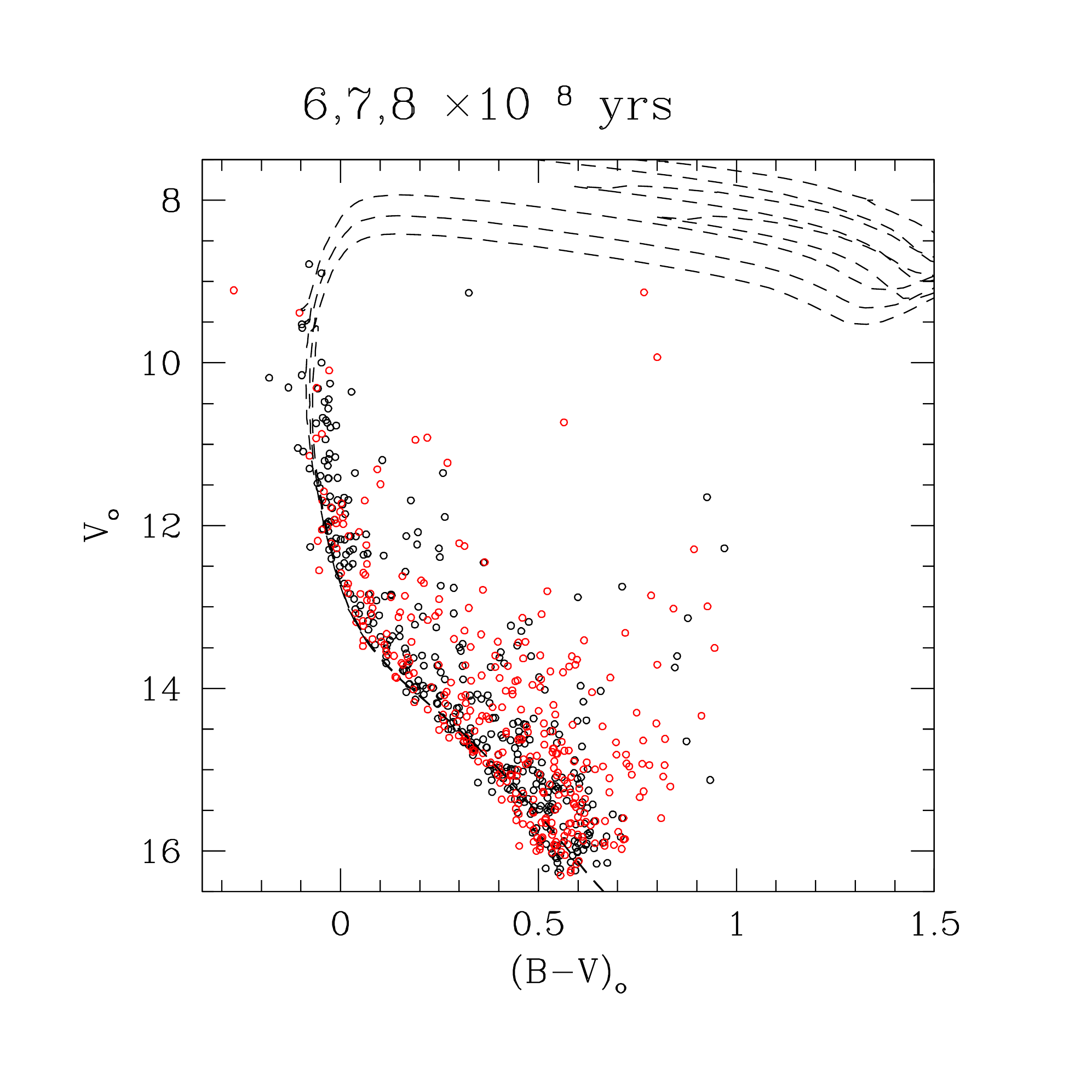}
\caption{Isochrone solution for NGC~5617 (black symbols) and Trumpler~22 (red symbols) stars. Isochrones are for ages of 60,70, and 80 million years, as indicated in the top of the Figure.}
\label{f:phot2}
\end{figure}
\end{center}

To estimate the age of Trumpler~22, we constructed the reddening-corrected CMD for the star within 4 arcmin form the center in Fig.\ref{f:phot2}. In this diagram, Trumpler~22 stars are indicated with red circles. Black circles are stars from NGC~5617, corrected for reddening as well. One can readily notice that the two clusters lie nicely one on top of the other, implying
they also share the same age. To illustrate this fact, three isochrones are shown, for ages of 6,7, and 8 $\times 10^8$ year, extracted from the PARSEC stellar models \citep{parsec} of solar metallicity. The fit supports an age around 70 Myr for both the clusters.

\section{Radial velocities}\label{sec:rv}
Spectra were collected on the 3.9m Anglo-Australian Telescope (AAT) using the UCLES spectrograph \citep{ucles} and using the HERMES multi-object spectrograph \citep{hermes} under service observation time. Summary of the spectroscopic observations are given in Table \ref{t:obs}. The star ID numbers for UCLES data are from \cite{Haug78} for NGC 5617 and from \cite{lindoff68} for Trumpler 22. The star ID numbers for HERMES data are from the photometry discussed in Section \ref{sec:phot} above.

With UCLES, we obtained spectra over 5-6 March 2012 for 6 bright cluster stars with a resolution of R $\sim$ 30,000 covering a wavelength range from $\sim$4550 \hbox{\AA} to $\sim$7200 \hbox{\AA}, using the $31$ l mm$^{-1}$ grating, the blue-sensitive EEV detector, a slit width of $1$\farcs$5$, and 2x binning in the spatial direction at a central wavelength of $5600$\hbox{\AA}. The typical signal to noise ratio at the central wavelength was 100 per pixel. The data were reduced using IRAF standard routines to subtract the bias level, apply a flat-field correction, identify and extract spectra, apply a wavelength calibration and co-add individual spectra. 

With HERMES, we obtained spectra on the 22 August 2014 for 50 stars from both clusters at a resolution of R $\sim$ 28,000 covering four wavelength ranges from 4710-4900\AA\  (Blue), 5650 - 5880\AA\ (Green), 6480 - 6740\AA\ (Red) and 7580 - 7890\AA\ (IR). Note that given the large field of view and multi-object capability of HERMES, a single pointing was sufficient to gather spectra of stars in both clusters. The typical signal to noise of the HERMES spectra were 100 per resolution element. The data were reduced using the 2dfdr\footnote{www.aao.gov.au/science/software/2dfdr} automatic data reduction pipeline dedicated to reducing multi-fibre spectroscopy data. 

\begin{table}
\begin{center}
\caption{Spectroscopic observations}
\label{t:obs}
\scriptsize
\begin{tabular}{c c c c c c c}
\hline\hline
Cluster & ID & R.A. & Dec & V & B-V & RV (\kms) \\
\hline
& & & UCLES\\
\hline
Trumpler 22 &32 &14 31 29.8 & -61 09 28.2 & 10.50 & 1.57 & -37.33 \\
Trumpler 22 &11 &14 31 09.5 & -61 09 43.0 & 10.7 & 1.30 & -38.96 \\
Trumpler 22 &49 &14 30 28.7 & -61 12 38.8 & 10.90 & 1.61 & -40.58 \\
NGC 5617 &283 &14 30 03.6 & -60 50 58.6 &  8.81 & 1.22 & -29.55 \\
NGC 5617 &116 &14 29 28.9& -60 42 18.90 & 10.67 & 1.75 & -36.85 \\
NGC 5617 &227 &14 29 50.5 & -60 41 13.1 &  10.42 & 1.64 & -35.20 \\
\hline
& & & HERMES\\
\hline                
Trumpler 22 &1145 &14 31 02.35 & -61 06 30.48 & 13.45 & 0.46 &  -54.42\\
Trumpler 22 &  911 &14 31 15.73 & -61 07 22.16 & 13.63 & 0.50 & -34.44\\
Trumpler 22 &  983 &14 31 11.17 & -61 09 11.95 & 12.69 & 0.37 & -39.03\\
Trumpler 22 &1079 &14 31 05.77 & -61 07 29.24 &  13.38 & 0.45 & -39.10\\
Trumpler 22 &1327 &14 30 52.44 & -61 07 11.06 &  14.00 & 0.81 & -1.62\\
Trumpler 22 &  989 &14 31 10.60 & -61 05 41.31 &  13.04 & 0.55 & -26.75\\
Trumpler 22 &  838 &14 31 20.21 & -61 10 15.60 &  13.48 & 0.44 & -39.97\\
Trumpler 22 &  799 &14 31 22.16 & -61 05 24.99 &  12.40 & 1.65 & -11.76\\
Trumpler 22 &1188 &14 31 00.53 & -61 11 44.39 &  13.24 & 0.40 & -37.11\\
Trumpler 22 &1136 &14 31 03.06 & -61 03 28.81 &  13.69 & 0.51 &-33.14\\
Trumpler 22 &1085 &14 31 05.47 & -61 10 04.24 &  12.47 & 0.67 & 37.92\\
Trumpler 22 &  794 &14 31 23.27 & -61 11 25.25 &  13.52 & 0.44 & -38.31\\
Trumpler 22 &1026 &14 31 08.89 & -61 13 09.48 &  13.07 & 0.45 & -38.73\\
Trumpler 22 &1122 &14 31 03.77 & -61 13 38.16 &  12.18 & 0.38 &-39.56\\
Trumpler 22 &1230 &14 30 58.09 & -61 13 07.24 &  11.72 & 0.42 &-39.80\\
Trumpler 22 &1417 &14 30 46.59 & -61 10 03.73 &  13.66 & 1.46 &-24.00\\
Trumpler 22 &1315 &14 30 53.23 & -61 12 35.99 &  13.83 & 0.44 &-39.81\\
Trumpler 22 &1222 &14 30 58.45 & -61 09 07.34 &  13.74 & 0.39 &-38.20\\
Trumpler 22 &1050 &14 31 07.20 & -61 08 32.12 &  11.86 & 0.39 &-39.34\\
Trumpler 22 &1363 &14 30 49.42 & -61 08 29.30 &  13.60 & 0.40 &-38.91\\
Trumpler 22 &1404 &14 30 47.00 & -61 07 24.17 &  12.28 & 1.01 &-36.47\\
Trumpler 22 &1381 &14 30 48.36 & -61 06 28.04 &  13.84 & 1.34 &-41.97\\
NGC 5617 &2517 &14 29 38.50 & -60 47 16.70 &  13.68  & 0.63 & -36.84\\
NGC 5617 &2748 &14 29 30.70 & -60 42 24.10 &  13.89  & 0.38 & -40.99\\
NGC 5617 &2930 &14 29 24.70 & -60 43 03.50 &  13.74  & 0.45 & -39.13\\
NGC 5617 &2498 &14 29 38.80 & -60 43 40.80 &  13.86 & 0.39 & -42.36\\
NGC 5617 &2392 &14 29 42.50 & -60 45 17.10 &  11.39 & 0.34 & -41.84\\
NGC 5617 &2148 &14 29 50.00 & -60 44 37.10 &  13.25 & 0.40 & -39.29\\
NGC 5617 &2950 &14 29 24.00 & -60 41 49.90 &  13.18 & 0.36 &-17.04\\
NGC 5617 &2822 &14 29 28.50 & -60 44 41.40 &  12.51 & 0.36 &-37.65\\
NGC 5617 &2382 &14 29 42.80 & -60 46 19.70 &  12.81 & 0.35 & -38.86\\
NGC 5617 &2699 &14 29 32.60 & -60 46 33.30 &  13.04 & 1.31 & -24.18\\
NGC 5617 &2550 &14 29 37.50 & -60 41 40.40 &  13.57 & 0.40 & -39.68\\
NGC 5617 &2348 &14 29 43.70 & -60 42 46.80 &  12.13 & 0.35 & -38.48\\
NGC 5617 &2180 &14 29 48.90 & -60 39 27.30 &  12.81 & 0.38 & -39.94\\
NGC 5617 &2619 &14 29 35.20 & -60 38 55.80 &  13.50 & 0.43 & -38.17\\
NGC 5617 &2181 &14 29 48.70 & -60 38 37.60 &  12.10 & 0.34 & -36.84\\
NGC 5617 &2078 &14 29 51.70 & -60 38 08.90 &  10.08 & 0.10 & -35.01\\
NGC 5617 &1728 &14 30 02.00 & -60 40 26.50 &  13.96 & 0.55 & -39.07\\
NGC 5617 &1935 &14 29 55.50 & -60 42 26.40 &  13.75&  0.40 & -39.65 \\
NGC 5617 &2191 &14 29 48.50 & -60 42 58.20 &  10.96& 0.30 & -36.80\\
NGC 5617 &2386 &14 29 42.60 & -60 41 51.30 &  12.70& 0.31 & -34.19\\
NGC 5617 &1687 &14 30 03.60 & -60 41 09.90 &  13.38& 0.35 & -43.71\\
NGC 5617 &1720 &14 30 02.40 & -60 42 20.50 &  12.48& 0.29 & -40.03\\
NGC 5617 &1762 &14 30 01.10 & -60 45 58.30 &  11.65& 0.36 & -40.54\\
NGC 5617 &1611 &14 30 06.10 & -60 43 16.90 &  13.14& 0.39 & -38.75\\
NGC 5617 &2202 &14 29 48.30 & -60 45 57.50 &  11.71& 1.84 & -15.86\\
NGC 5617 &1912 &14 29 56.30 & -60 44 35.00 &  10.53& 0.71 & -39.37\\
NGC 5617 &2074 &14 29 52.00 & -60 43 30.30 &  13.52& 0.55 & -36.25\\
NGC 5617 &2135 &14 29 50.30 & -60 47 27.70 &  13.63& 0.58 & -37.75\\
\hline\hline
\end{tabular}
\end{center}
\end{table}

\begin{table}
\caption{Cluster radial velocities}
\begin{center}
\begin{tabular}{ccc}
\hline\hline                    
& Trumpler 22	 & NGC 5716 \\
\hline
$<$RV$>$ $\pm$ std (\kms) & -38.46 $\pm$ 2.08 &  -38.63 $\pm$ 2.25 		\\
\hline\hline
\end{tabular}
\end{center}
\label{t:rv}
\end{table}

The radial velocities (RV) for the UCLES spectra were measured using the IRAF\footnote{IRAF is distributed by the National Optical Astronomy Observatories, which are operated by the Association of Universities for Research in Astronomy, Inc., under cooperative agreement with the National Science Foundation.} $fxcor$ package, using a solar reference template. For HERMES spectra, correlating against a solar template produced large errors due to differences in spectral type between the Sun and the cluster sample. Instead the HERMES spectra were cross correlated against an internal reference spectrum, Trumpler 22 star \#1122. Careful manual measurement of the spectral features in star \#1122 found it has a RV = -39.56 $\pm$ 1.11 \kms. The derived relative RVs of the remaining HERMES spectra were converted to absolute values by applying the RV of the reference star \#1122. Note that of the four channels in HERMES, the IR channel wavelength is contaminated by the atmospheric A-band absorption features. For this reason, the IR channel was not used for measuring RVs. The typical uncertainty in the RV measurements across the remaining three HERMES channels is of the order of 1-2 \kms. The typical uncertainty in RVs for the UCLES spectra are less than 1 \kms. 
%We suspect the channel to channel differences in HERMES spectra are most likely driven by issues with the automatic %wavelength calibration routines. 

The final heliocentric corrected RVs of the cluster stars are given in Table \ref{t:obs}. The RV analysis reveals that 6 stars in Trumper 22 (star \#1145, \#1327, \#989, \#799, \#1085 and \#1417) and 4 stars in NGC 5617 (star \#2950, \#2699, \#2202 and \#283) have different RVs compared to the majority of the cluster sample stars. These stars are most likely to be non-members of the respective clusters and are probably field stars that happen to lie within the field of view (see also Section \ref{sec:chem} below). Disregarding these stars, the average RV for both clusters are summarised in Table \ref{t:rv}. The two clusters have nearly identical RVs, and the measured velocity dispersion in both clusters are comparable with open cluster dispersions. The measured RV for NGC 5617 sample is in agreement with the measurements of \cite{frinchaboy} and \cite{Mermilliod}, who report RV values between -35.77 to -36.60 \kms, when taking into account the error budget. We have two stars, \#116 and \#227, in common with both these studies. The difference between us and \cite{Mermilliod} is $\delta$RV = 0.04 \kms and 0.22 \kms for these two stars respectively. The difference between us and \cite{frinchaboy} is $\delta$RV = -0.19 \kms and 0.52 \kms for these two stars respectively. In all cases the difference in RV is well within the errors.

\section{Chemical abundances}\label{sec:chem}

Of the total spectra collected, the majority of the targets are either too hot or too fast rotators, such that there are either no measurable spectral line features or all features have been smoothed out. A high fraction of hot stars and rapid rotators are expected in young clusters such as the Trumpler 22 and NGC 5617. Most of the stars that show some measurable spectral features are mostly non-members of the open clusters' as determined by the RV analysis.

%We note several interesting features in the spectra. Several stars (ID \#838 in Trumpler 22, and IDs \#2392, 2498, 2135, %1762, 2148 and 1611 in NGC 5617) show H emission features, most likely due to chromospheric activity, debris disks {\bf(??)} in these young objects. In particular stars \#2392 and \#1762 in NGC 5617 contain strong H$_{\alpha}$ and H$_{\beta}$ emission. In some cases the emission profiles are seen only in H$_{\alpha}$ only. {\bf (Why??)} Of the spectra that %display absorption features beyond hydrogen, several stars have large Fe features, most notably the Fe I line at 6677 \AA\ %and 6613 \AA\, particularly large in stars \#1122 ,1404 in Trumper 22 and star \# 2191, 1912 in NGC 5617. {\bf Why? What %are these stars?}. None of the spectra displayed any double lined binaries.

\begin{center}
\begin{figure}
\includegraphics[width=9cm]{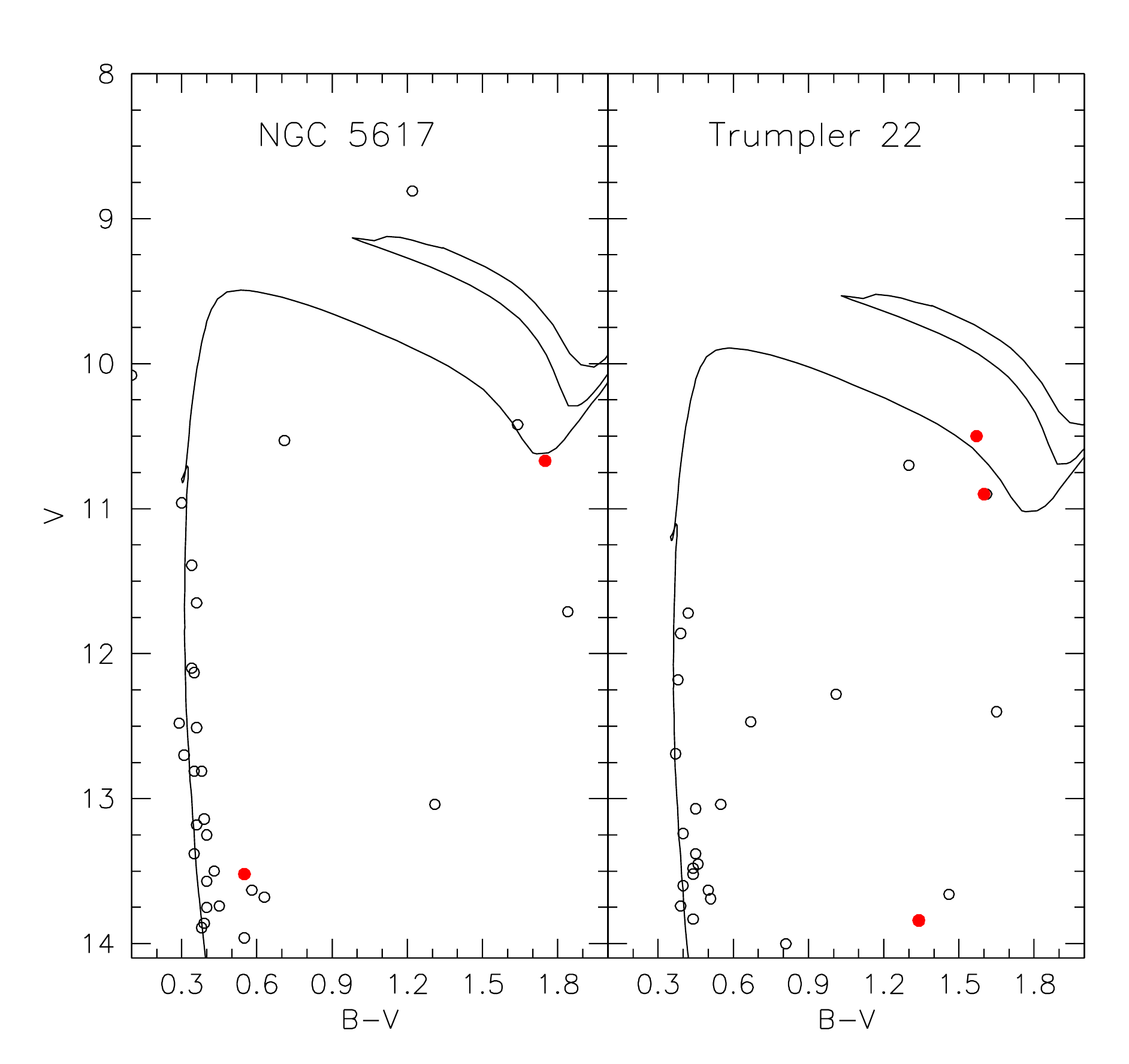}
\caption{Colour-magnitude diagrams of NGC~5617 and Trumpler~22 for the spectroscopic sample stars, with the stars subject to spectroscopic analysis highlighted in filled red circles. Overlaid is the isochrone for 70 Myr.}
\label{f:cmd}
\end{figure}
\end{center}

We carried out a spectroscopic analysis of all RV member stars that displayed a sufficient number of measurable lines, which consists of 3 stars in Trumpler 22 and 2 stars in NGC 5617. The spectroscopic sample and their derived stellar parameters are given in Table \ref{t:params}. Their location on the cluster colour-magnitude diagram are shown in Figure \ref{f:cmd}. Based on its location on the colour-magnitude diagram and the derived parameters, Trumpler 22 stars \#1381 appears to be a non-member, despite having a similar RV and metallicity as the cluster stars.

For stellar parameters and abundance determinations we use the MOOG code \citep{moog} and interpolated Kurucz model atmospheres based on the ATLAS9 code \citep{Castelli97} with no convective overshooting. The stellar parameter determination procedure was as follows. The equivalent width (EW) of Fe  {\sc i} and Fe  {\sc ii} lines were measured from the spectra. Estimates of the effective temperature (\teff) and \logg\ for each target were made using photometric calibrations of \cite{casagrande10} to use as a first guess model. Effective temperature (\teff) was derived by requiring excitation equilibrium of the Fe~{\sc i} lines.  Micro-turbulence was derived from the condition that abundances from Fe {\sc i} lines show no trend with EW. Surface gravity (\logg) was derived via ionisation equilibrium, i.e. requiring the abundances from Fe {\sc i} equal Fe {\sc ii}. The number of spectral lines used in the analysis varied due to the difference in wavelength covered by the UCLES and HERMES spectrographs. At minimum 20 Fe {\sc i} lines and 3 Fe {\sc ii} lines were used in the analysis of HERMES spectra, while the UCLES spectra covered $\sim$50 Fe {\sc i} lines and 5-10 Fe {\sc ii} lines. The uncertainty on the derived stellar parameters is of the order $\delta$ \teff\ = 50K, $\delta$ \logg\ = 0.1 dex, $\delta$ micro-turbulence = 0.1 \kms and $\delta$ [Fe/H] = 0.1 dex. 

Using the derived stellar parameters the abundances for Na, Mg, Al, Si, Ca and Ni were measured based on line EW measurements. The error on the derived abundances were estimated by quadratically summing the uncertainty in the measured EWs together with the errors due to stellar parameters, which were estimated by varying one parameter at a time, and checking the corresponding variation in the resulting abundances. The total error per star per element is given in Table \ref{t:results}.

The measured abundances find that most elements are in their solar proportions, with the exception of Na. Na abundances are most likely affect by NLTE effects. We make use of the INSPECT web tool at www.inspect-stars.net to determine the impact of NLTE effects on the Na abundances \citep{lind2011}. We find the NLTE correction is of the order of -0.15 dex, which brings to mean Na abundance in the clusters to slightly above solar level at [Na/Fe] $\sim$ 0.18 dex. 

%The results presented in Table \ref{t:params} and \ref{t:results} show that both clusters share similar chemical properties, with $<[Fe/H]>$ = -0.18$\pm$0.02 dex as derived from the RV members. This is consistent with the common motion of the stars, suggesting that the two clusters are co-natal. 

\begin{table}
\caption{Stellar parameters}
\begin{center}
\begin{tabular}{c c c c c c c c c c c c c c c c c}
\hline\hline                    
Cluster	 & ID & T$_{\rm eff}$ & log$g$ & $\xi$ & $[$Fe/H$]$ \\
\hline
Trumpler 22 &   32 & 4555 & 1.6 & 2.72 & -0.15 \\
Trumpler 22 &   49 & 4970 & 2.1 & 2.78 & -0.19 \\
Trumpler 22 &1381& 4600 & 1.5 & 1.35 & -0.16 \\
\hline
NGC 5617 &116   &4450 & 1.2 & 2.44 & -0.20 \\
NGC 5617 &2074 & 5800& 3.9 &1.20  & -0.15\\
\hline\hline
\end{tabular}
\end{center}
\label{t:params}
\end{table}

\begin{table*}
\caption{Chemical abundances}
\begin{threeparttable}
\begin{center}
\begin{tabular}{c c c c c c c c c c c c c c c c c}
\hline\hline                    
Cluster	 & ID & $[$Na/Fe$]$ & $[$Mg/Fe$]$ & $[$Al/Fe$]$ & $[$Si/Fe$]$ & $[$Ca/Fe$]$ & $[$Ni/Fe$]$  \\
\hline
Trumpler 22 &   32 & 0.36$\pm$0.06 & -0.02$\pm$0.07& 0.02$\pm$0.05 & 0.02$\pm$0.07 & -0.07$\pm$0.05 & 0.03$\pm$0.07\\
Trumpler 22 &   49 & 0.39$\pm$0.05 & 0.09$\pm$0.05 & 0.08$\pm$0.06 & 0.00$\pm$0.05 & -0.06$\pm$0.07 & 0.06$\pm$0.05 \\
Trumpler 22 &1381& 0.34$\pm$0.15 & 0.05$\pm$0.06 & 0.13$\pm$0.10 & 0.03$\pm$0.15 & 0.03$\pm$0.15  & 0.05$\pm$0.11 \\
\hline
Average \tnote{a}       &        &  0.38$\pm$0.02& 0.04$\pm$0.08& 0.05$\pm$0.04& 0.01$\pm$0.01&-0.07$\pm$0.01&0.05$\pm$0.02 \\
\hline
NGC 5617 &116   & 0.30$\pm$0.07 & -0.05$\pm$0.06 & 0.05$\pm$0.06 & 0.05$\pm$0.09  & -0.03$\pm$0.08 & 0.02$\pm$0.08\\
NGC 5617 &2074 & 0.21$\pm$0.04 & -0.02$\pm$0.05 & -0.01$\pm$0.05 & 0.00$\pm$0.05 & -0.05$\pm$0.06 &0.01$\pm$0.05 \\
\hline
Average     &        & 0.26$\pm$0.06& -0.04$\pm$0.02& 0.02$\pm$0.04& 0.03$\pm$0.04 & -0.04$\pm$0.01&0.02$\pm$0.01\\
\hline\hline
\end{tabular}
\begin{tablenotes}
        \item[a] Average does not include star 1381.
    \end{tablenotes}
\end{center}
\end{threeparttable}
\label{t:results}
\end{table*}

\section{Summary and conclusions}\label{sec:summary}
In this paper, we explored the properties of the open cluster NGC 5617 and Trumpler 22. We presented new CCD photometry, from which we find the two clusters share a common age of 70$\pm$10 Myrs and a common distance of 2.1$\pm$0.3 kpc. We examined their RVs based on high resolution spectra with the AAT-UCLES and AAO-HERMES spectrographs and present RV members of the two clusters. The two clusters share a common average RV of $\sim$38.5$\pm$2.0 \kms. While the bulk of the spectra collected were either too hot or too fast rotators to carry out spectroscopic analysis, we identified 3 members in Trumpler 22 and 2 members in NGC 5617 suitable for chemical analysis. However star \#1381 in Trumpler 22 is most likely to be a non-member based on its position on the CMD, although it has a RV and metallicity similar to that of the cluster members. The abundance results found that both clusters share a common chemical enrichment history with [Fe/H] = -0.18 $\pm$ 0.02 dex and no significant differences were seen in the other studied elements between the two clusters within the measurement errors.

The presented evidence is consistent with the two clusters being co-natal, confirming that our Galaxy is a host for primordial binary clusters. Further exploration of a larger sample of binary clusters candidates is encouraged to determine the primordial binary cluster fraction within the Galaxy.

\section*{Acknowledgments}
We thank the anonymous referee for helpful suggestions which have improved the paper. G. Carraro acknowledges financial support from the AAO Distinguished Visitor program and ESO DGDF program during a visit at AAO where part of this work was done.
We thank the Centre de Donn\'{e}es Astronomiques de Strasbourg (CDS), the U. S. Naval Observatory and NASA for the use of their electronic facilities, especially SIMBAD, ViZier and ADS. This paper has made extensive use of the Webda database at http://www.univie.ac.at/webda.

\bibliography{References}
%\label{lastpage}

%\begin{thebibliography}{64}
%\expandafter\ifx\csname natexlab\endcsname\relax\def\natexlab#1{#1}\fi

%\end{thebibliography}

%\bibliographystyle{mnras} % style aa.bst
%\bibliography{mybib} % your references Yourfile.bib

\appendix
\section[]{Atomic line list}\label{appendix:lines}
The full table of atomic line data is available online.
\begin{center}
%\begin{supertable}
%\caption{Atomic line lists}
\begin{supertabular}{lcccccccccr}
\hline\hline
Element & Wavelength (\AA) & LEP (eV) & log $gf$ & EW (m\AA)  \\
\hline
\multicolumn{4}{c}{Trumpler 22: \#32 } \\
\hline
Fe\,{\sc i}	&	4808.15	&	3.25	&	-2.79	&	89	\\
Fe\,{\sc i}	&	5067.15	&	4.22	&	-0.97	&	153	\\
Fe\,{\sc i}	&	5217.39	&	3.21	&	-1.07	&	192	\\
Fe\,{\sc i}	&	5285.13	&	4.43	&	-1.64	&	86	\\
Fe\,{\sc i}	&	5322.04	&	2.28	&	-2.8	&	188	\\
Fe\,{\sc i}	&	5373.71	&	4.47	&	-0.76	&	132	\\

\hline\hline
\end{supertabular}
%\end{supertable}
\end{center}

\end{document}